\documentclass{IEEEtran}
\usepackage{cite}
\usepackage{amsmath,amssymb,amsfonts}
\usepackage{algorithmic}
\usepackage{graphicx}
\usepackage{textcomp}
\usepackage{xcolor}
\usepackage{mathrsfs}
\usepackage{psfrag}
\usepackage{cancel}
\usepackage{verbatim}
\usepackage{setspace}

\usepackage{gensymb}

\usepackage{circuitikz}
\usetikzlibrary{decorations.pathreplacing,calligraphy,patterns}

\makeatletter
\let\NAT@parse\undefined
\makeatother
\usepackage{hyperref}
\usepackage{theoremref}

\def\mb{\mathbb}

\def\BibTeX{{\rm B\kern-.05em{\sc i\kern-.025em b}\kern-.08em
    T\kern-.1667em\lower.7ex\hbox{E}\kern-.125emX}}

\begin{document}
\sloppy
\title{\LARGE \bf Discrete-time Integral Resonant Control of Negative Imaginary Systems: Application to a High-speed Nanopositioner}
\author{{Kanghong Shi},$\quad${Erfan Khodabakhshi},$\quad${Prosanto Biswas}, $\quad${Ian R. Petersen, \IEEEmembership{Life Fellow, IEEE}},\\ and {S. O. Reza Moheimani, \IEEEmembership{Fellow, IEEE}}
\thanks{This work was supported by the Australian Research Council under grant DP230102443, and partially by the UTD Center for Atomically Precise Fabrication of Solid-state Quantum Devices.}
\thanks{Kanghong Shi and Ian R. Petersen are with the School of Engineering, College of Engineering, Computing and Cybernetics, Australian National University, Canberra, ACT 2601, Australia. Erfan Khodabakhshi, Prosanto Biswas and S. O. Reza Moheimani are with the Erik Jonsson School of Engineering and Computer Science, The University of Texas at Dallas, Richardson, TX 75080 USA. Corresponding author: S. O. Reza Moheimani.
{\tt kanghong.shi@anu.edu.au}, {\tt erfan@utdallas.edu}, {\tt prosanto.biswas@utdallas.edu}, {\tt ian.petersen@anu.edu.au}, {\tt reza.moheimani@utdallas.edu}}}

\newtheorem{definition}{Definition}
\newtheorem{theorem}{Theorem}
\newtheorem{conjecture}{Conjecture}
\newtheorem{lemma}{Lemma}
\newtheorem{remark}{Remark}
\newtheorem{corollary}{Corollary}
\newtheorem{assumption}{Assumption}

\maketitle

\thispagestyle{empty}
\pagestyle{empty}

\begin{abstract}
We propose a discrete-time integral resonant control (IRC) approach for negative imaginary (NI) systems, which overcomes several limitations of continuous-time IRC. We show that a discrete-time IRC has a step-advanced negative imaginary property. A zero-order hold-sampled NI system can be asymptotically stabilized using a discrete-time IRC with suitable parameters. A hardware experiment is conducted where a high-speed flexure-guided nanopositioner is efficiently damped using the proposed discrete-time IRC with the discrete-time controller being implemented in FPGA hardware at the sampling rate of $\mathbf{1.25\,\mathrm{MHz}}$.
\end{abstract}


\section{INTRODUCTION}

Control design for highly resonant flexible structures poses significant challenges since their structural modes can limit the stability margin and impair the performance of the controlled system \cite{pota1999resonant,devasia2007survey,nikooienejad2021frequency}. For instance, in atomic force microscopy (AFM), the resonant mode of the positioner negatively impacts the maximum closed-loop bandwidth, restricting the AFM to only low-speed scanning \cite{devasia2007survey,moheimani2008invited,habibullah2019robust}. Additionally, external disturbances can trigger the resonance, further undermining closed-loop performance \cite{fleming2009new,eielsen2013damping}. Therefore, it is essential to effectively dampen these resonant modes to simplify the control system design and ensure the closed-loop system's stability and robustness \cite{nikooienejad2021frequency}.


Those equipped with collocated actuator-sensor pairs in lightly damped flexible structures can facilitate control design, especially when dealing with plant uncertainties and unmodeled dynamics. \cite{petersen2011negative,bhikkaji2011negative}. A system with collocated force actuators and position sensors typically has the negative imaginary (NI) property \cite{lanzon2008stability,petersen2010feedback}. Based on the NI stability theorem \cite{lanzon2008stability}, the positive feedback interconnection of an NI plant $G(s)$ and a strictly negative imaginary (SNI) controller $R(s)$ will be internally stable, provided that the largest eigenvalue of the DC loop gain is less than unity; i.e., $\lambda_{\max}(G(0)R(0))<1$ \cite{lanzon2008stability}.

NI systems theory was introduced in \cite{lanzon2008stability,petersen2010feedback} to address the robust control problem for flexible structures \cite{preumont2018vibration,halim2001spatial,pota2002resonant}. Several difficulties are associated with the control of flexible structures, such as variable resonance frequencies, high system order, and highly resonant dynamics. These challenges can severely compromise the performance of the control system or cause instability if the controller is not robust against these uncertainties \cite{omidi2014hybrid,petersen2016negative,khodabakhshi2022characterization,feng2023high}. NI systems theory offers a framework to evaluate the robustness and design robust controllers for flexible structures in the case of collocated force actuators and position sensors \cite{lanzon2008stability,petersen2010feedback,petersen2016negative}.

Roughly speaking, a square, real-rational, proper transfer matrix $G(s)$ is said to be NI if it is stable and for all positive frequencies $\omega>0$, its frequency response $G(j\omega)$ has negative imaginary parts. An NI system can be regarded as the cascade of a passive system with an integrator. In comparison to passivity theory, which can deal with systems of relative degree zero and one \cite{brogliato2007dissipative}, an advantage of NI systems theory is that it can deal with systems of relative degree zero, one and two \cite{shi2024necessary,dannatt2023strictly}. Since it was introduced, NI systems theory has attracted attention among control theorists (see e.g., \cite{xiong2010negative,mabrok2014generalizing,wang2015robust,ghallab2018extending,shi2023output,bhowmick2017lti,mabrok2015generalized}) and has found its application in many fields including nanopositioning control \cite{mabrok2013spectral,das2014mimo,das2015multivariable,shi2023MEMS,nikooienejad2021convex}, the control of lightly damped structures \cite{cai2010stability,rahman2015design,bhikkaji2011negative}, and the control of electrical power systems \cite{chen2023design,chen2023nonlinear}, etc.


Several popular SNI controllers, including the positive position feedback (PPF) controller \cite{russell2017evaluating,GC85,FC90,C05e,J06d,J14a}, the resonant controller (RC) \cite{pota2002resonant,halim2001spatial,J05d,J08g}, and the integral resonant controller (IRC) \cite{aphale2007integral,J08h,fleming2009new,pereira2010integral}, have been implemented successfully. These controllers are not only easy to implement but also highly effective in terms of providing robust damping, especially when dealing with uncertainties in resonance frequency \cite{russell2017evaluating}. Among these methods, integral resonant control stands out as a straightforward, low-order approach that can damp multiple modes while maintaining stability margins \cite{russell2017evaluating,nikooienejad2021frequency,khodabakhshi2022characterization}.



The general concept of an integral resonant controller is to modify the pole-zero interlacing of a collocated system $G(s)$ into a zero-pole interlacing pattern by: a) introducing a feed-through term $D$ to the system $G(s)$ and, b) adding an integral controller $C(s) = \Gamma/s$ to the augmented system $\widehat{G}(s) = G(s) + D$; see \cite{aphale2007integral,yong2008design}. When the integral gain $\Gamma$ increases, the system's poles shift away from the imaginary axis and deeper into the left-half complex plane, finally aligning with the open-loop zero locations. Owing to its straightforward implementation and robust performance, IRC has been widely adopted to enhance damping in a variety of applications, including cantilever beams \cite{aphale2007integral,bhikkaji2008multivariable,maclean2020modified}, flexible robotic manipulators \cite{pereira2010integral}, nanopositioning stages \cite{khodabakhshi2022characterization}, atomic force microscopes \cite{wadikhaye2014control,yong2015collocated}, flexible civil structures \cite{basu2011multi,beskhyrounintegral}, and floors subject to vibrations induced by walking \cite{beskhyrounintegral}.

Most literature on nanopositioning focuses on control design within the continuous $s-$domain rather than the discrete-time $z-$domain. These control strategies are typically articulated through continuous-time state-space models. For real-time applications, however, it is necessary to discretize these models and solve them in discrete-time steps using a fixed step size. While this discretization process enables the practical application of continuous-time controllers in real-time scenarios, transforming a continuous-time design into a discrete-time implementable version may introduce inaccuracies, particularly for designing damping controllers at higher frequencies. 
Such a control design method, where a controller is designed based on a continuous-time model and then discretized, is called the continuous-time design or CTD; see \cite{nesic1999sufficient} and the references therein. A drawback of the CTD technique is its dependency on a sampling rate that is high enough to ensure stability. However, attaining such a high sampling frequency can exceed the capabilities of some devices \cite{owens1990fast}. Also, as for the digital control of a nanopositioner, the continuous-time model of a nanopositioner is usually estimated via sampled data, which introduces inaccuracies to the control design. To overcome these limitations of a continuous-time IRC, we seek to construct a discrete-time IRC that is designed according to an accurate model of the plant and allows for a relatively slower sampling rate.

In this article, we introduce a discrete-time IRC for the digital control of real-world NI systems. We consider the discrete-time NI system property introduced in \cite{shi2023discrete}, where the discrete-time NI property will always be satisfied by a zero-order hold (ZOH) sampled continuous-time NI system (see \cite{aastrom2013computer} for a discussion of ZOH sampling). Under some assumptions, a discrete-time NI system can be stabilized using another discrete-time NI system whose output takes a one-step advance, which is called a discrete-time step advanced negative imaginary (SANI) system. We show that a discrete-time IRC is an SANI system. Also, we show that the interconnection of a linear NI system and a discrete-time IRC is asymptotically stable if the IRC parameters satisfy certain conditions related to the DC gain of the plant. Therefore, the digital control problem for an NI system reduces to the process of finding a pair of suitable parameters based on the DC gain of the NI plant. Also, since the NI property is usually guaranteed by a system's physical nature and is preserved under ZOH sampling, the plant's model does not need to be known in the control design process, yet the closed-loop stability is still rigorously guaranteed.

Compared with the continuous-time IRC, the advantages of using a discrete-time IRC are two-fold: a) The controller design process is more straightforward as we do not need to reconstruct a continuous-time model of the plant and discretize the controller. This significantly saves the computational resources required in generating the control input; b) Stability is rigorously guaranteed and less reliant on a high sampling rate since the stability analysis is carried out purely in discrete time without any type of approximation.

When implementing a discrete-time controller, selecting an appropriate step size is crucial to maintain stability. If the step size is too large, the solver may become unstable, necessitating a reduction in step size. However, there is a practical lower limit on how small the step size can be, especially when the controller runs on hardware with limited computational capabilities. The processing unit requires adequate time to execute necessary calculations at each step. For instance, controlling a system with a high mechanical bandwidth demands a high control loop bandwidth, requiring a smaller step size to ensure accurate implementation. If the step size is inadequate, the controller might not function as intended, compromising system performance. Therefore, ensuring that the processing speed is sufficient to handle the controller's computational complexity within the bandwidth of the system is vital.

The discrete-time IRC is designed and implemented for a high-speed, high-bandwidth nanopositioning system. The controller implementation was performed using a LabVIEW program and a National Instruments PXIe-7975R FlexRIO module. A NI-5782 adapter module with a $250\,\mathrm{MHz}$ clock speed was used to acquire the digital data and implement the discrete-time IRC at a sampling rate of $1.25\,\mathrm{MHz}$.

The rest of this article is organized as follows: Section \ref{sec:pre} provides preliminary results on continuous-time IRC and also discrete-time NI systems theory. In Section \ref{section:DT_IRC}, we introduce the discrete-time IRC. To be specific, we provide the system model of a discrete-time IRC and show that it is an SANI system. We also prove that the interconnection of a discrete-time NI system and a discrete-time IRC is asymptotically stable under some conditions on the IRC parameters. In Section \ref{sec:Experiment}, a hardware experiment is conducted where a discrete-time IRC is applied to a high-speed flexure-guided nanopositioner. We conclude the article in Section \ref{sec:conclusion}.

Notation: The notation in this paper is standard. $\mathbb R$ denotes the field of real numbers. $\mb N$ denotes the set of nonnegative integers. $\mathbb R^{m\times n}$ denotes the space of real matrices of dimension $m\times n$. $A^T$ denotes the transpose of a matrix $A$.  $A^{-T}$ denotes the transpose of the inverse of $A$; that is, $A^{-T}=(A^{-1})^T=(A^T)^{-1}$. $\lambda_{max}(A)$ denotes the largest eigenvalue of a matrix $A$ with real spectrum. $\|\cdot\|$ denotes the standard Euclidean norm. For a real symmetric or complex Hermitian matrix $P$, $P>0\ (P\geq 0)$ denotes the positive (semi-)definiteness of a matrix $P$ and $P<0\ (P\leq 0)$ denotes the negative (semi-)definiteness of a matrix $P$. A function $V: \mb R^n \to \mb R$ is said to be positive definite if $V(0)=0$ and $V(x)>0$ for all $x\neq 0$.
\section{PRELIMINARIES}\label{sec:pre}
\subsection{Continuous-time integral resonant controller}
\begin{figure}[h!]
\centering
\psfrag{r}{$r$}
\psfrag{e_s}{$e(s)$}
\psfrag{U_s}{$U(s)$}
\psfrag{Y_s}{$Y(s)$}
\psfrag{barY_s}{$\overline Y(s)$}
\psfrag{G_s}{$G(s)$}
\psfrag{C_s}{$C(s)$}
\psfrag{D}{$D$}
\includegraphics[width=8.5cm]{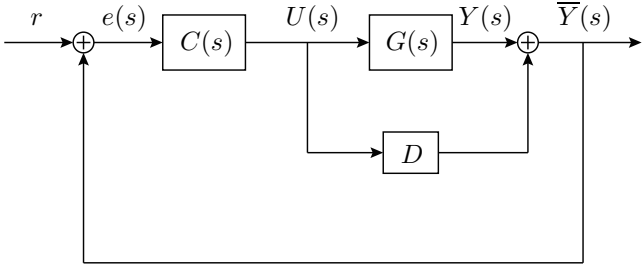}
\caption{Closed-loop interconnection of an integrator $C(s)=\frac{\Gamma}{s}$ and $G(s)+D$.}
\label{fig:CT_IRC}
\end{figure}

\begin{figure}[h!]
\centering
\psfrag{r}{$r$}
\psfrag{e_s}{$e(s)$}
\psfrag{U_s}{$U(s)$}
\psfrag{Y_s}{$Y(s)$}
\psfrag{G_s}{$G(s)$}
\psfrag{K_s}{$K(s)$}
\includegraphics[width=8.5cm]{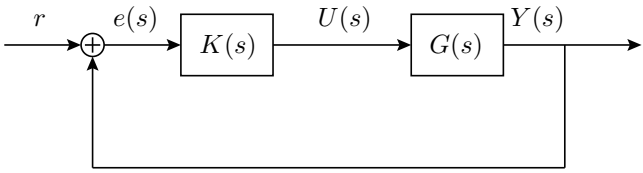}
\caption{Closed-loop interconnection of an IRC and a plant. This is equivalent to the closed-loop system in Fig.~\ref{fig:CT_IRC}.}

\label{fig:CT_IRC_equivalence}
\end{figure}

Integral resonant control (IRC) was introduced in \cite{aphale2007integral,bhikkaji2008multivariable} to provide damping for flexible structures with collocated and compatible actuator/sensor pairs (e.g. force actuators and position sensors). Fig.~\ref{fig:CT_IRC} shows how a continuous-time IRC is implemented. Given a plant with a square transfer matrix $G(s)$, we apply a direct feedthrough $D$ and also an integral controller
\begin{equation}\label{eq:CT integrator}
	C(s)=\frac{\Gamma}{s}
\end{equation}
in positive feedback with $G(s)+D$. Here, we require the matrices $\Gamma, D\in \mb R^{p\times p}$ to satisfy $D<0$ and $\Gamma>0$. The relation between the plant input $U(s)$ and the plant output $Y(s)$ is described as follows:
\begin{equation}\label{eq:U(s) Y(s) relation}
	U(s)= C(s)e(s)
	=C(s)\left(r+Y(s)+DU(s)\right).
\end{equation}
Hence,
\begin{equation}\label{eq:U(s) in terms of Y(s) and C(s)}
	U(s)=(I-C(s) D)^{-1}C(s)(r+Y(s)).
\end{equation}
According to (\ref{eq:U(s) in terms of Y(s) and C(s)}), the closed-loop system shown in Fig.~\ref{fig:CT_IRC} can be equivalently constructed as the interconnection of $G(s)$ and $K(s)$ shown in Fig.~\ref{fig:CT_IRC_equivalence}, where
\begin{equation}\label{eq:K(s) in terms of C(s)}
	K(s):=(I-C(s) D)^{-1}C(s).
\end{equation}
Substituting (\ref{eq:CT integrator}) into (\ref{eq:K(s) in terms of C(s)}) yields
\begin{equation}\label{eq:CT IRC}
	K(s):=\left(sI-\Gamma D\right)^{-1}\Gamma.
\end{equation}
Here, $K(s)$ given in (\ref{eq:CT IRC}) is the transfer matrix of a continuous-time IRC. An IRC is an SNI system, and can be used in the control of NI plants (see \cite{bhikkaji2008multivariable,lanzon2008stability,petersen2010feedback}).

\subsection{Discrete-time NI systems}\label{subsec:pre_NI}
We consider the definition of discrete-time NI systems introduced in \cite{shi2023discrete}. This property is automatically satisfied by any ZOH sampled continuous-time NI system. A discrete-time NI system can be stabilized by another discrete-time NI system with a step advance. First, we provide the definition of discrete-time NI systems given in \cite{shi2023discrete}.

Consider the system
\begin{subequations}\label{eq:DT_nonlinear}
\begin{align}
	x_{k+1} =&\ f(x_k,u_k),\label{eq:state eq}\\
	y_k=&\ h(x_k),\label{eq:output eq}
\end{align}	
\end{subequations}
where $f:\mathbb R^n \times \mathbb R^p\to \mathbb R^n$ and $h:\mathbb R^n \to \mathbb R^p$. Here $u_k,y_k \in \mathbb R^p$ and $x_k\in \mathbb R^n$ are the input, output and state of the system at time step $k\in \mathbb N$, respectively.

\begin{definition}[discrete-time negative imaginary systems]\label{def:DT_NNI}\cite{shi2023discrete}
The system (\ref{eq:DT_nonlinear}) is said to be a discrete-time negative imaginary (NI) system if there exists a continuous positive definite function $V\colon \mb R^n \to \mb R$ such that for arbitrary $x_k$ and $u_k$,
\begin{equation}\label{eq:NNI ineq}
V(x_{k+1})-V(x_{k})\leq u_k^T\left(y_{k+1}-y_{k}\right),	
\end{equation}
for all $k$.
\end{definition}

In the case of a linear system, necessary and sufficient conditions are given under which the system satisfies Definition \ref{def:DT_NNI}. Consider a linear system of the form
\begin{subequations}\label{eq:G(z)}
	\begin{align}
\Sigma\colon\ 		x_{k+1} =&\ Ax_k+Bu_k,\label{eq:G(z) state eq}\\
		y_k =&\ Cx_k,
	\end{align}
\end{subequations}
where $x_k\in \mathbb R^n$, $u_k,y_k\in \mb R^p$ are the system state, input and output, respectively. Here $A\in \mb R^{n\times n}$, $B\in \mb R^{n\times p}$ and $C\in \mb R^{p\times n}$.
\begin{lemma}\label{lemma:LMI DT-NI}\cite{shi2023discrete}
Suppose the linear system (\ref{eq:G(z)}) satisfies $\det(I-A)\neq 0$. Then the system (\ref{eq:G(z)}) is NI with a quadratic positive definite storage function satisfying (\ref{eq:NNI ineq}) if and only if there exists a real matrix $P=P^T>0$ such that
\begin{equation}\label{eq:LMIs in NI lemma}
	A^TPA-P\leq 0 \quad \textnormal{and} \quad C = B^T(I-A)^{-T}P.
\end{equation}
\end{lemma}

A discrete-time NI system can be stabilized using a discrete-time step advanced NI (SANI) system. An SANI system can be obtained by replacing the output of an NI system using the output at the next time step.
We provide the definition of SANI systems in the following. Consider the system 
\begin{subequations}\label{eq:nonlinear SANI system}
\begin{align}
\widetilde x_{k+1} =&\ \widetilde f(\widetilde x_k, \widetilde u_k),\\
	\widetilde y_k=&\ \widetilde h(\widetilde x_k, \widetilde u_k),
\end{align}	
\end{subequations}
where $\widetilde f:\mathbb R^n \times \mathbb R^p\to \mathbb R^n$ and $\widetilde h:\mathbb R^n \to \mathbb R^p$. Here $\widetilde u, \widetilde y \in \mathbb R^p$ and $\widetilde x\in \mathbb R^n$ are the input, output and state of the system at time step $k\in \mathbb N$, respectively.
\begin{definition}\label{def:SANI}\cite{shi2023discrete}
	The system \eqref{eq:nonlinear SANI system} is said to be a step-advanced negative imaginary (SANI) system if there exists a function $\widehat h(x_k)$ such that:
	\begin{enumerate}
		\item $\widetilde h(\widetilde x_k, \widetilde u_k)=\widehat h(\widetilde f(\widetilde x_k, \widetilde u_k))$;
		\item There exists a continuous positive definite function $\widetilde V\colon \mb R^n \to \mb R$ such that for arbitrary state $\widetilde x_k$ and input $\widetilde u_k$,
\begin{equation*}
\widetilde V(\widetilde x_{k+1})-\widetilde V(\widetilde x_{k})\leq \widetilde u_k^T\left(\widehat h(\widetilde x_{k+1})-\widehat h(\widetilde x_{k})\right)
\end{equation*}
for all $k$.
	\end{enumerate}
\end{definition}
\begin{remark}\label{remark:NI_SANI}
	Definition \ref{def:SANI} can be regarded as a variant of Definition \ref{def:DT_NNI} in a way such that the system output $y_k$ takes the next step output value; i.e., $h(x_{k+1})$. To be specific, suppose the system (\ref{eq:DT_nonlinear}) is NI as per Definition \ref{def:DT_NNI}. Then a system with the same state equation (\ref{eq:state eq}) and an output equation $\widetilde y(k) = h(x(k+1))=h(f(x_k,u_k))$ is an SANI system. Note that this does not affect the causality of the system because $h(f(x_k,u_k))$ is a function of the state $x_k$ and input $u_k$ of the current step $k$.
\end{remark}

\section{DISCRETE-TIME INTEGRAL RESONANT CONTROLLER}\label{section:DT_IRC}
Since real-world control systems are usually implemented digitally using computers \cite{aastrom2013computer}, we aim to find the discrete-time version of the IRC in order to provide digital control for NI plants. Considering that sampling a continuous-time NI plant yields a discrete-time NI system, a discrete-time IRC is required to serve as a controller for a discrete-time NI system.

As is mentioned in Section \ref{subsec:pre_NI}, a discrete-time NI system can be stabilized using a discrete-time SANI system. A continuous-time NI system $K(s)$, which has SNI property, is expected to become a discrete-time NI system after ZOH sampling. Therefore, taking a step advance of a ZOH sampled IRC will result in an SANI system. However, a disadvantage of taking the ZOH discretization of $K(s)$ is that the model of the resulting system will include matrix exponential terms. In order to achieve a neat system model for the discrete-time IRC, we first construct a discrete-time system of the similar form as $K(s)$ in (\ref{eq:K(s) in terms of C(s)}), except that the continuous-time integrator $C(s)$ is replaced by a discrete-time integrator
\begin{equation}
	C(z)=\frac{\Gamma}{z-1}.
\end{equation}
This discrete-time system has the transfer matrix
\begin{align}
	K(z)=&\left(I-C(z)D\right)^{-1}C(z)\notag\\
	=&\left(I-\frac{\Gamma D}{z-1}\right)\frac{\Gamma}{z-1}\notag\\
	=&\left[zI-(I+\Gamma D)\right]^{-1}\Gamma.\label{eq:DT K(z)}
\end{align}
We still require that $\Gamma>0$ and $D<0$. However, the matrices $\Gamma$ and $D$ here are not related to those in the continuous-time IRC. A state-space realization of the transfer matrix $K(z)$ is given as follows:
\begin{subequations}\label{eq:K(z) realization}
	\begin{align}
		\widetilde x_{k+1}=&\ (I+\Gamma D)\widetilde x_k+\Gamma \widetilde u_k,\\
		\widehat y_k=&\ \widetilde x_k,
	\end{align}
\end{subequations}
where $\widetilde x_k,\widetilde u_k,\widehat y_k\in \mb R^p$ are system state, input and output, respectively. We show in the following that the system (\ref{eq:K(z) realization}) is a discrete-time NI system when certain conditions are satisfied for the matrices $\Gamma$ and $D$.

\begin{lemma}\label{lemma:K(z) NI}
	Suppose $\Gamma>0$ and $-2\Gamma^{-1}\leq D<0$, then the system (\ref{eq:K(z) realization}) with the transfer matrix $K(z)$ is a discrete-time NI system.
\end{lemma}
\begin{IEEEproof}
	We apply Lemma \ref{lemma:LMI DT-NI} to show the NI property of (\ref{eq:K(z) realization}). Let
	\begin{align}
		\overline A =&\ (I+\Gamma D),\\
		\overline B =&\ \Gamma,\\
		\overline C =&\ I.
	\end{align}
Then $\det(I-\overline A)=\det(-\Gamma D)\neq 0$. We show in the following that the matrix $\overline P = -D$ satisfies the conditions in (\ref{eq:LMIs in NI lemma}). We have

\begin{align}
	\overline A^T\overline PA-\overline P=&-D-2D\Gamma D-D\Gamma D\Gamma D +D\notag\\
	=&-2D\Gamma D-D\Gamma D\Gamma D\notag\\
	=& -D\Gamma\left(2\Gamma^{-1}+D\right)\Gamma D\notag\\
	\leq &\ 0.
\end{align}
Also, we have
\begin{align}
	\overline B^T(I-\overline A)^{-T}\overline P=\Gamma(-\Gamma D)^{-T}(-D)=I=\overline C.
\end{align}
Therefore, the system (\ref{eq:K(z) realization}) with the transfer matrix $K(z)$ in (\ref{eq:DT K(z)}) is NI.
\end{IEEEproof}

We have shown that the system (\ref{eq:K(z) realization}), which is constructed in a similar way as the continuous-time IRC, is a discrete-time NI system. However, according to the control framework for discrete-time NI systems in \cite{shi2023discrete}, an NI system can be applied as the controller for another NI system after introducing one step advance in its output equation. Therefore, we introduce one step advance to the output of the system (\ref{eq:K(z) realization}), which yields the following system.
\begin{subequations}\label{eq:DT IRC model}
	\begin{align}
		\widetilde x_{k+1}=&\ (I+\Gamma D)\widetilde x_k+\Gamma \widetilde u_k,\\
		\widetilde y_k=&\ (I+\Gamma D)\widetilde x_k+\Gamma \widetilde u_k,
	\end{align}
\end{subequations}
where $\widetilde x_k,\widetilde u_k,\widehat y_k\in \mb R^p$ are system state, input and output, respectively. Here, $\Gamma>0$ and $-2\Gamma^{-1}\leq D<0$. The transfer matrix of the system (\ref{eq:DT IRC model}) is
\begin{equation}
F(z)=z\left[zI-(I+\Gamma D)\right]^{-1}\Gamma.
\end{equation}
We call the system (\ref{eq:DT IRC model}) the discrete-time integral resonant controller. Since the system is obtained by taking one step advance of the NI system (\ref{eq:K(z) realization}), it is an SANI system and hence we expect the interconnection of a linear NI plant with the system (\ref{eq:DT IRC model}) to be at least Lyapunov stable. We provide a particular stability proof in the following and show that asymptotic stability can indeed be achieved for the interconnection of an NI plant and a discrete-time IRC of the form (\ref{eq:DT IRC model}). First, we prove the SANI property of the discrete-time IRC (\ref{eq:DT IRC model}).
\begin{lemma}\label{lemma:DT IRC SANI}
	Suppose $\Gamma>0$ and $-2\Gamma^{-1}\leq D<0$. Then the discrete-time integral resonant controller (\ref{eq:DT IRC model}) is an SANI system.
\end{lemma}
\begin{IEEEproof}
	The proof follows directly from Definitions \ref{def:DT_NNI} and \ref{def:SANI} and Lemma \ref{lemma:K(z) NI}. To be specific, since the system (\ref{eq:DT IRC model}) is obtained by taking one step advance of the NI system (\ref{eq:K(z) realization}), then the function $\widehat h(x_k)=x_k$ satisfies Condition 1) in Definition \ref{def:SANI} and Condition 2) is satisfied due to the NI property of the system (\ref{eq:K(z) realization}) as shown in Lemma \ref{lemma:K(z) NI}.
\end{IEEEproof}

Now consider a linear discrete-time system with the minimal realization
\begin{subequations}\label{eq:plant}
	\begin{align}
		x_{k+1}=&\ Ax_k+Bu_k,\\
		y_k=&\ Cx_k,
	\end{align}
\end{subequations}
where $x_k\in \mb R^n$, $u_k,y_k\in \mb R^p$ are the system state, input and output, respectively. Here, $A\in \mb R^{n\times n}$, $B\in\mb R^{n\times p}$ and $C\in \mb R^{p\times n}$. The transfer matrix of the system (\ref{eq:plant}) is
\begin{equation}
	G(z)=C(zI-A)^{-1}B.
\end{equation}
We show in the following that if the system (\ref{eq:plant}) is NI, then there exists a discrete-time IRC of the form (\ref{eq:DT IRC model}) such that their closed-loop interconnection is asymptotically stable.

\begin{figure}[h!]
\centering
\psfrag{r}{\hspace{-0.2cm}$r=0$}
\psfrag{u_c}{$\widetilde u_k$}
\psfrag{y_c}{$\widetilde y_k$}
\psfrag{u_p}{$u_k$}
\psfrag{y_p}{$y_k$}
\psfrag{F_z}{$F(z)$}
\psfrag{G_z}{$G(z)$}
\includegraphics[width=8.5cm]{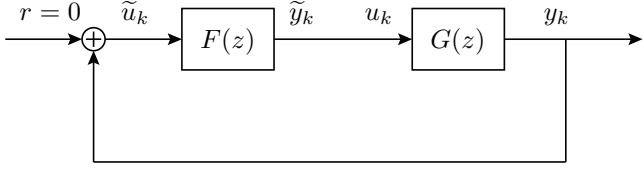}
\caption{Positive feedback interconnection of the plant $G(z)$ of the form (\ref{eq:plant}) and a discrete-time IRC $F(z)$ of the form (\ref{eq:DT IRC model}).}
\label{fig:closed-loop interconnection of a plant and a IRC}
\end{figure}

\begin{theorem}
Consider the closed-loop interconnection of the minimal system (\ref{eq:plant}) and a discrete-time IRC of the form (\ref{eq:DT IRC model}), as is shown in Fig.~\ref{fig:closed-loop interconnection of a plant and a IRC}. Suppose the system (\ref{eq:plant}) with the transfer matrix $G(z)$ satisfies $\det(I-A)\neq 0$ and is an NI system according to Definition \ref{def:DT_NNI}. If the discrete-time IRC (\ref{eq:DT IRC model}) satisfies $-2\Gamma^{-1}< D<-G(1)$, then the closed-loop system shown in Fig.~\ref{fig:closed-loop interconnection of a plant and a IRC} is asymptotically stable. 
\end{theorem} 
\begin{IEEEproof}
	According to Lemma \ref{lemma:LMI DT-NI}, the NI property of the plant (\ref{eq:plant}) implies that there exists a symmetric positive definite matrix $P\in \mb R^{n\times n}$ such that
	\begin{align}
		A^TPA-P\leq  0;\label{eq:APA-P}\\
		C = B^T(I-A)^{-T}P.\label{eq:C=B(I-A)P}
	\end{align}
According to the system setting $\widetilde u_k=y_k$ and $u_k=\widetilde y_k$ as shown in Fig.~\ref{fig:closed-loop interconnection of a plant and a IRC}, we have
\begin{align}
	x_{k+1}=&\ Ax_k+Bu_k\notag\\
	=&\ Ax_k+B\widetilde y_k\notag\\
	=&\ Ax_k+B[(I+\Gamma D)\widetilde x_k+\Gamma \widetilde u_k]\notag\\
	=&\ Ax_k+B[(I+\Gamma D)\widetilde x_k+\Gamma y_k]\notag\\
	=&\ Ax_k+B[(I+\Gamma D)\widetilde x_k+\Gamma C x_k]\notag\\
	=&\ (A+B\Gamma C) x_k + B(I+\Gamma D)\widetilde x_k,\label{eq:x_{k+1} in closed-loop}
\end{align}
and
\begin{align}
	\widetilde x_{k+1}=&\ (I+\Gamma D)\widetilde x_k +\Gamma \widetilde u_k\notag\\
	=&\ (I+\Gamma D)\widetilde x_k +\Gamma y_k\notag\\
	=&\ (I+\Gamma D)\widetilde x_k +\Gamma C x_k,\label{eq:tilde x_{k+1} in closed-loop}
\end{align}
where the equalities also use (\ref{eq:DT IRC model}) and (\ref{eq:plant}). Therefore, according to (\ref{eq:x_{k+1} in closed-loop}) and (\ref{eq:tilde x_{k+1} in closed-loop}), the closed-loop system shown in Fig.~\ref{fig:closed-loop interconnection of a plant and a IRC} has the state-space model
\begin{equation}
	\begin{bmatrix}
		x_{k+1}\\ \widetilde x_{k+1}
	\end{bmatrix}=\begin{bmatrix}
		A+B\Gamma C & B+B\Gamma D\\ \Gamma C & I+\Gamma D
	\end{bmatrix}\begin{bmatrix}
		x_k\\ \widetilde x_k
	\end{bmatrix}.
\end{equation}
Let
\begin{equation}
	\widehat A := \begin{bmatrix}
		A+B\Gamma C & B+B\Gamma D\\ \Gamma C & I+\Gamma D
	\end{bmatrix}.
\end{equation}
In what follows, we apply Lyapunov's stability theorem. We construct the candidate Lyapunov function of the closed-loop system in Fig.~\ref{fig:closed-loop interconnection of a plant and a IRC} to be:
\begin{equation}
	W(x_k,x_{k+1})=\frac{1}{2}\begin{bmatrix}
		x_k \\ \widetilde x_k
	\end{bmatrix}^T\begin{bmatrix}
P & -C^T \\ -C & -D\end{bmatrix}\begin{bmatrix}
		x_k \\ \widetilde x_k
	\end{bmatrix},
\end{equation}
where $P$ is the positive definite matrix that satisfies (\ref{eq:APA-P}) and (\ref{eq:C=B(I-A)P}). Let
\begin{equation}
	Q := \begin{bmatrix}
P & -C^T \\ -C & -D\end{bmatrix}.
\end{equation}
We prove in the following that $Q>0$. The condition (\ref{eq:C=B(I-A)P}) implies
\begin{equation}\label{eq:B=(I-A)PC}
	B=(I-A)P^{-1}C^T.
\end{equation}
Using (\ref{eq:B=(I-A)PC}), we have that
\begin{equation}\label{eq:G(1)>0}
	G(1)=C(I-A)^{-1}B=CP^{-1}C^T>0.
\end{equation}
Since $P>0$ and the Schur complement of the block $P$ of the matrix $Q$ is
\begin{equation}
	Q/P=-D-CP^{-1}C^T=-D-G(1),
\end{equation}
which is positive definite according to the assumption on $D$, then we have $Q>0$. Since $Q>0$ and $D<0$, we also have that
\begin{equation}\label{eq:P+CDC}
	Q/(-D)=P+C^TDC>0.
\end{equation}

Now we take the increment of the candidate Lyapunov function $W(x_k,\widetilde x_k)$ to show stability. We have
\begin{align}
	W(x_{k+1},\widetilde x_{k+1})&-W(x_k,\widetilde x_k)\notag\\
	=&\ \frac{1}{2}\begin{bmatrix}
		x_k \\ \widetilde x_k
	\end{bmatrix}^T\left(\widehat A^TQ\widehat A-Q\right)\begin{bmatrix}
		x_k \\ \widetilde x_k
	\end{bmatrix}.\label{eq:W increment}
\end{align}
Using (\ref{eq:B=(I-A)PC}), we have that
\begin{equation}
	\widehat A^TQ\widehat A-Q=\begin{bmatrix}
		M_{11}&M_{12}\\M_{12}^T & M_{22}
	\end{bmatrix},
\end{equation}
where
\begin{align}
	M_{11}=&\ \left(C^T\Gamma C P^{-1}-I\right)\left(A^TPA-P\right)\left(I-P^{-1}C^T\Gamma C\right)\notag\\
	&-2C^T\Gamma C-C^T\Gamma D\Gamma C;\label{eq:M_11}\\
	M_{12}=&\ \left(C^T\Gamma C P^{-1}-I\right)\left(A^TPA-P\right)P^{-1}C^T(I+\Gamma D)\notag\\
	&-C^T\Gamma D(2I+\Gamma D);\label{eq:M_12}\\
	M_{22}=&\ (I+D\Gamma)CP^{-1}\left(A^TPA-P\right)P^{-1}C^T(I+\Gamma D)\notag\\
	&-D\Gamma D\Gamma D-2D\Gamma D.\label{eq:M_22}
\end{align}
According to (\ref{eq:M_11})--(\ref{eq:M_22}), we have that
\begin{align}
	\widehat A^T & Q\widehat A-Q\notag\\
	=& \begin{bmatrix}
		C^T\Gamma C-P\\(I+D\Gamma)C
	\end{bmatrix}P^{-1}\left(A^TPA-P\right)P^{-1}\begin{bmatrix}
		C^T\Gamma C-P\\(I+D\Gamma)C
	\end{bmatrix}^T\notag\\
	&-\begin{bmatrix}
		C^T\Gamma \\ D\Gamma
	\end{bmatrix}\left(D+2\Gamma^{-1}\right)\begin{bmatrix}
		C^T\Gamma \\ D\Gamma
	\end{bmatrix}^T.\label{eq:hat AQhat A-Q in two  matrices}
\end{align}
Here, we have $A^TPA-P\leq 0$ according to (\ref{eq:APA-P}) and $D+2\Gamma^{-1}> 0$ according to the assumption of the theorem. Therefore, both terms on the right-hand side (RHS) of (\ref{eq:hat AQhat A-Q in two  matrices}) are negative semidefinite. Hence, the matrix $\widehat A^T Q\widehat A-Q$ is negative semidefinite. This implies that $W(x_{k+1},\widetilde x_{k+1})-W(x_k,\widetilde x_k)\leq 0$.
That is, the closed-loop system shown in Fig.~\ref{fig:closed-loop interconnection of a plant and a IRC} is stable in the sense of Lyapunov. We will show in the following that the closed-loop system is asymptotically stable using LaSalle's invariance principle (see e.g., \cite{haddad2008nonlinear,mei2017lasalle}). According to (\ref{eq:W increment}) and (\ref{eq:hat AQhat A-Q in two  matrices}), if $W(x_{k+1},\widetilde x_{k+1})-W(x_k,\widetilde x_k)$ stays at zero, then the scalar

\begin{equation}
\begin{bmatrix}
	x_k\\ \widetilde x_k
\end{bmatrix}^T
	\begin{bmatrix}
		C^T\Gamma \\ D\Gamma
	\end{bmatrix}\left(D+2\Gamma^{-1}\right)\begin{bmatrix}
		C^T\Gamma \\ D\Gamma
	\end{bmatrix}^T\begin{bmatrix}
	x_k\\ \widetilde x_k
\end{bmatrix}\label{eq:second term in AQA-Q}
\end{equation}
also stays at zero. This is because both terms on the RHS of (\ref{eq:hat AQhat A-Q in two  matrices}) are negative semidefinite and can only contribute non-positive values to $W(x_{k+1},\widetilde x_{k+1})-W(x_k,\widetilde x_k)$. If (\ref{eq:second term in AQA-Q}) stays at zero, then since $\left(D+2\Gamma^{-1}\right)>0$ and $\Gamma>0$, it follows that
\begin{equation}
	\begin{bmatrix}
		C & D
	\end{bmatrix}\begin{bmatrix}
	x_k\\ \widetilde x_k
\end{bmatrix}
\end{equation}
stays at zero. That is $Cx_k+D\widetilde x_k\equiv 0$. In this case,
\begin{equation}\label{eq:tilde x_k in terms of x_k}
	\widetilde x_k\equiv -D^{-1}Cx_k.
\end{equation}
Substituting (\ref{eq:tilde x_k in terms of x_k}) in (\ref{eq:tilde x_{k+1} in closed-loop}), we have
\begin{equation}\label{eq:tilde x constant}
	\widetilde x_{k+1}\equiv \widetilde x_k
\end{equation}
Since $\det D \neq 0$, then (\ref{eq:tilde x_k in terms of x_k}) and (\ref{eq:tilde x constant}) together imply that for all future time steps $k$
\begin{equation}\label{eq:Cx_k constant}
	Cx_k\equiv \alpha,
\end{equation}
for some constant vector $\alpha$. According to the observability of the system (\ref{eq:plant}),  (\ref{eq:Cx_k constant}) implies
\begin{equation}
	x_{k+1}\equiv x_k.
\end{equation}
Substituting (\ref{eq:tilde x_k in terms of x_k}) in (\ref{eq:x_{k+1} in closed-loop}), we have
\begin{equation}\label{eq:x_{k+1} stays constant}
	x_{k+1}\equiv (A-BD^{-1}C)x_k.
\end{equation}
Using (\ref{eq:B=(I-A)PC}), we have that
\begin{align}
	A-BD^{-1}C=&\ A-(I-A)P^{-1}C^TD^{-1}C\notag\\
	=&\ (A-I)P^{-1}(P+C^TDC)+I.\label{eq:A-BDC}
\end{align}
We have $\det(A-I)\neq 0$ according to the assumption of the theorem and also $\det(P+C^TDC)\neq 0$ according to (\ref{eq:P+CDC}). Therefore,
\begin{equation}\label{eq:first term in A-BDC nonsingular}
	(A-I)P^{-1}(P+C^TDC)x_k\neq 0
\end{equation}
for any nonzero $x_k$. Substituting (\ref{eq:A-BDC}) in (\ref{eq:x_{k+1} stays constant}) with (\ref{eq:first term in A-BDC nonsingular}) also considered, we have that
\begin{equation}
	x_{k+1}\neq x_k
\end{equation}
unless $x_k=0$. In the case that $x_k=0$, we also have $\widetilde x_k=0$ according to (\ref{eq:tilde x_k in terms of x_k}). Hence, the system state already stays at the equilibrium at the origin. Otherwise, $W(x_{k+1},\widetilde x_{k+1})-W(x_k,\widetilde x_k)$ cannot stay at zero. Therefore, $W(x_k,\widetilde x_k)$ will keep decreasing until it reaches zero. That is, the system state also reaches the origin. Therefore, the closed-loop interconnection shown in Fig.~\ref{fig:closed-loop interconnection of a plant and a IRC} is asymptotically stable.
\end{IEEEproof}
\begin{remark}
	The condition $-2\Gamma^{-1}< D<-G(1)$ of the IRC parameters can be fulfilled in two steps. Given the matrix $G(1)$, we choose $D$ such that $D<-G(1)$. Then we choose $\Gamma$ such that $\Gamma <-2D^{-1}$. Note that since $G(1)>0$ according to (\ref{eq:G(1)>0}), then for matrices $\Gamma$ and $D$ that satisfy $-2\Gamma^{-1}< D<-G(1)$, they also satisfy the conditions in Lemma \ref{lemma:DT IRC SANI}.
\end{remark}

\section{HIGH-SPPED FLEXURE-GUIDED NANOPOSITIONER}\label{sec:Experiment}
In this section, a discrete-time IRC is designed and implemented on a high-speed flexure-guided nanopositioner as a linear NI system to demonstrate the applicability of the stability result.

\begin{figure}[t]
	\centering
	\includegraphics[width=8.5cm]{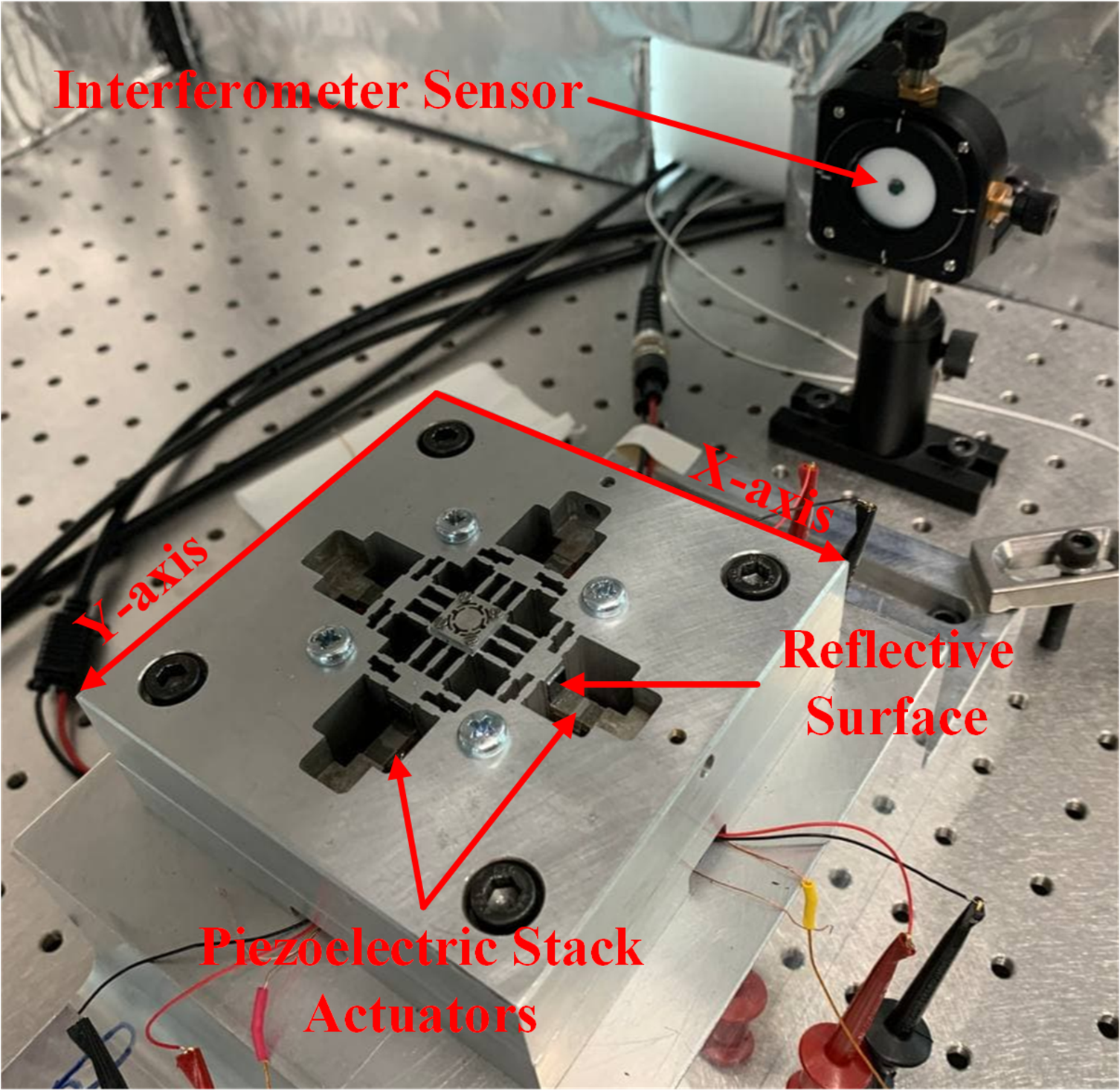}
	\caption{\footnotesize Nanopositioner and interferometer sensor}\label{fig:expsetup}
\end{figure}

\begin{figure}[t]
	\centering
	\includegraphics[width=8.5cm]{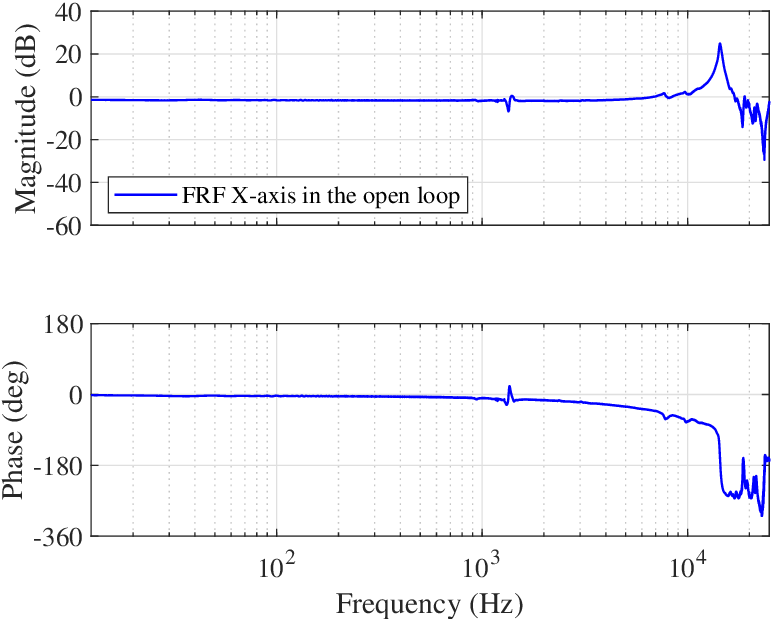}
	\caption{\footnotesize Frequency response of the nanopositioner.}\label{fig:frfx}
\end{figure}

Figure\ref{fig:expsetup} demonstrates the experimental setup of a high-speed flexure-guided nanopositioner introduced in \cite{khodabakhshi2022design}. The nanopositioning stage is driven by four PiezoDrive PX200 voltage amplifiers (each with a gain of 20) for all piezoelectric actuators. A PICOSCALE interferometer is used for high-precision displacement measurements of the nanopositioner in both $X-$ and $ Y-$directions. Under ideal conditions, the lightly-damped nanopositioner with collocated actuators and sensors would be an NI system. However, due to the fabrication tolerances and bandwidth limitations of the driving electronics, the nanopositioner violates the NI system property beyond a certain frequency where the phase surpasses $-180\degree$. Figure \ref{fig:frfx} shows the frequency response function (FRF) of the $X-$axis nanopositioner from actuation to the sensor output. The fundamental resonance frequency of the nanopositioner along $X-$axis is at $14.86\,\mathrm{kHz}$. We see that the phase drops beyond $15.01\,\mathrm{kHz}$, and the system violates the NI property as frequency increases. However, as the frequency response rolls off, the phase deviation is negligible. Therefore, the frequency response of the system can be approximated by a NI model within the frequency range of interest. The implementation of the designed IRC is described in the next sections.

\subsection{Experimental Setup}
\begin{figure}[bt]
    \centering
	\includegraphics[width=8.5cm]{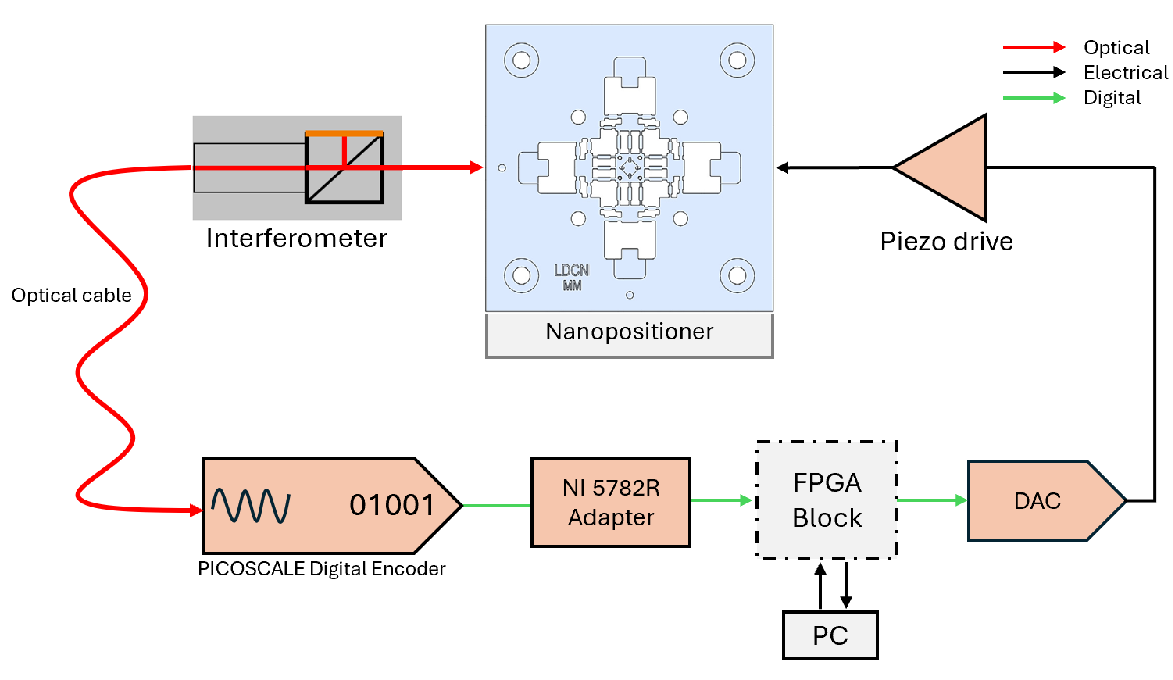}
    \caption{\footnotesize A schematic of the experimental apparatus.}\label{fig:setup}
\end{figure}

Figure (\ref{fig:setup}) shows a schematic representation of the experimental apparatus. A high-precision Michelson-interferometer is employed to measure the displacement of the scanner. The measurement from the optical sensor is transmitted as Digital Differential Interface (DDI) outputs via the PICOSCALE Breakout Box (BOB). These DDI outputs configured as a Quadrature (AquadB) signal from the BOB are then carefully adjusted for the desired resolution within our implementable bandwidth in the form of step size and step frequency. Subsequently, we implement a standard quadrature decoder within the LabVIEW FPGA environment to derive position values in $\mathrm{\mu m}$ alongside the designed discrete-time integral resonant controller.

\begin{figure}[ht]
  \centering
  \includegraphics[width=7.5cm]{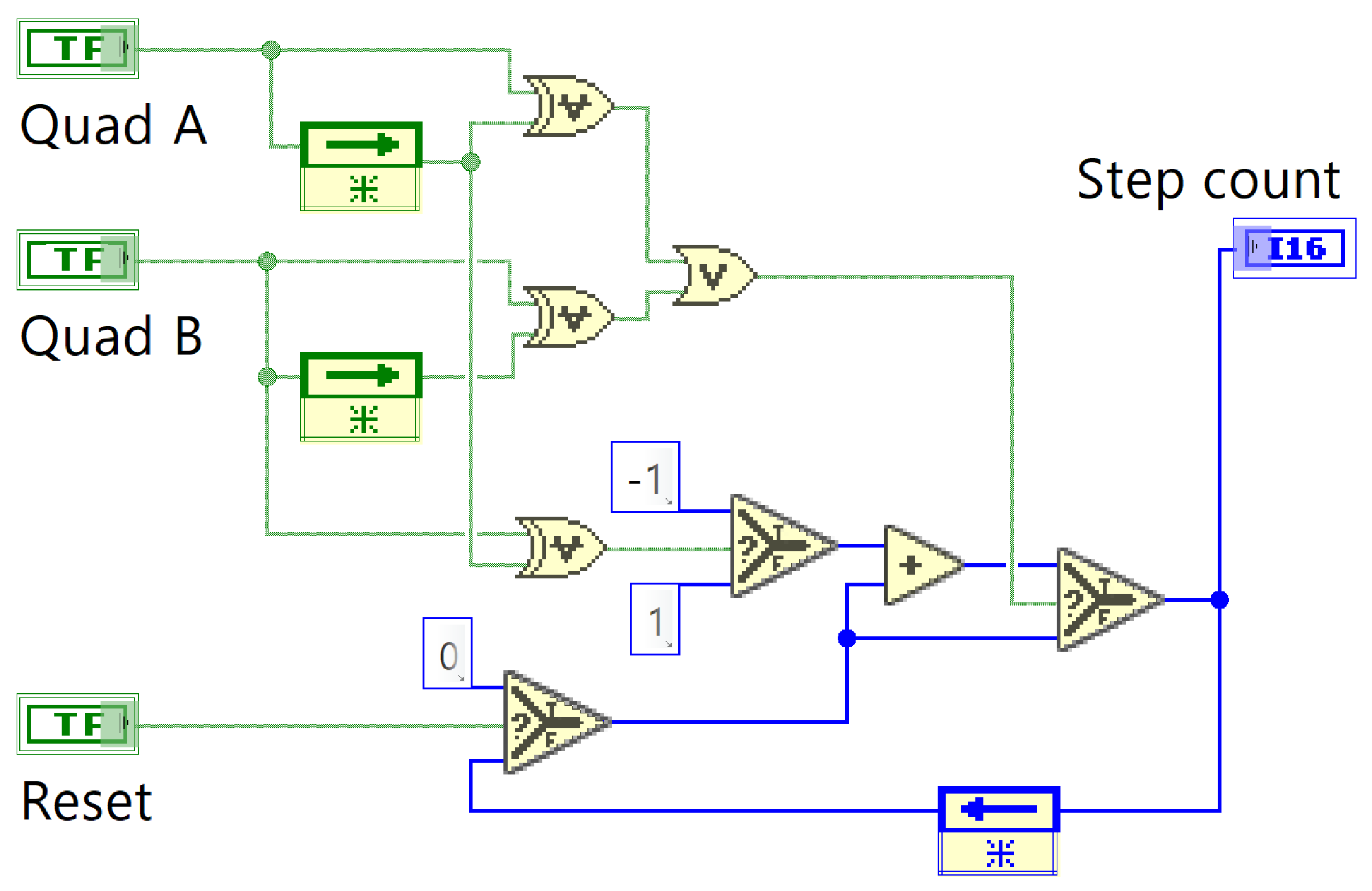}
  \caption{Quadrature decoder implemented in LabVIEW FPGA.}
  \label{fig:QDecoder}
\end{figure}

The operating principle of the quadrature decoder is as follows: it utilizes two phases of the encoder, Quad A and Quad B, which are spaced at 90-degree intervals. This arrangement allows the logic operations to determine the direction of movement. The decoder uses the current state of B and the previous state of A to obtain the direction of movement and adjust the counting value accordingly, either adding or subtracting one count from the previous value. Since PICOSCALE only measures relative position changes, an initialization process is necessary for accurate counting, facilitated by a reset circuit. Subsequently, each count is multiplied by a position scaling factor determined from the step size. Through experimentation, it is found that the sensor noise is slightly below $6\,\mathrm{nm}$, leading to the configuration of the sensor step size to be precisely $6\,\mathrm{nm}$ for each count. Figure \ref{fig:QDecoder} illustrates an excerpt of the implemented quadrature decoder in the LabVIEW environment, responsible for converting DDI output into position parameters.

\subsection{FPGA Implementation}
\begin{figure}[b]
	\centering
    \includegraphics[width=7cm]{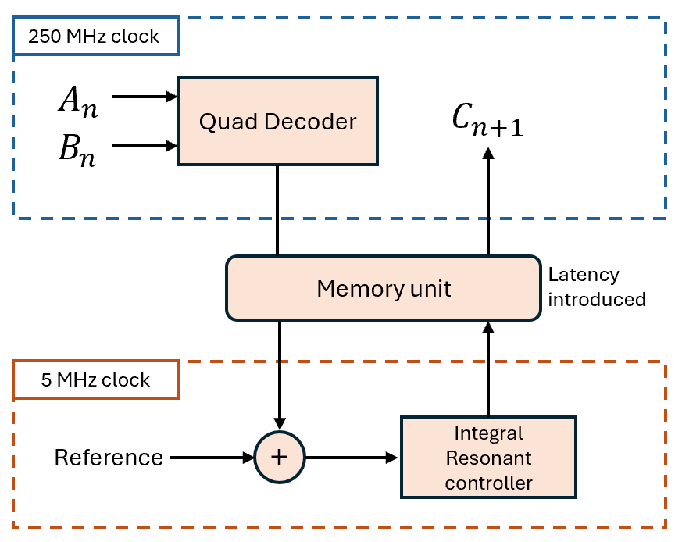}
    \caption{\footnotesize Parallel computation inside FPGA.}\label{fig:fpga}
\end{figure}

In this experiment, a Kintex-7 XC7K410T FPGA embedded in a National Instruments PXIe-7975R FlexRIO module is adopted as the core computing engine. This module interfaces with an adapter module, offering hardware description language integration capability. The FPGA hardware is developed using the graphical programming language in the LabVIEW FPGA environment. Connected to the FlexRIO module is the National Instruments 5782R adapter module, a 14-bit, four-channel digitizer with a bandwidth of $500\,\mathrm{MHz}$ and a sampling rate of $250\,\mathrm{MS/s}$. Each component is programmed within a single-cycled timed loop (SCTL) to optimize hardware design performance and minimize computational latency. This loop ensures specific timing cycles for executing one iteration, provided there is no timing violation during compilation. Additionally, a voltage converter is employed to adjust the DDI output level from the BOB to match the adapter's input voltage level. With its high sampling rate of $250\,\mathrm{MHz}$, the decoder swiftly transforms digital data into position parameters. However, the involvement of multiplications in our first-order discrete-time IRC within each loop, spanning multiple clock cycles, surpasses the $4\,\mathrm{ns}$ time limit. As a result, the decoded position data is transmitted to a slower SCTL, which operates at a clock frequency of $5\,\mathrm{MHz}$. This frequency is derived from the primary oscillator of $250\,\mathrm{MHz}$, as illustrated in figure (\ref{fig:fpga}). This additional data transfer latency, coupled with register usage between loops, further reduces the achievable sampling rate to $1.25\,\mathrm{MHz}$, necessitating the implementation of the discrete-time IRC at this reduced rate. 

When dealing with a high clock frequency, computational speed becomes an issue. Integer-type variables are favored due to their faster processing, resulting in shorter computation times, especially in the high-frequency loop. Conversely, in the slower loop, fixed-point data formats are employed for variables and parameters, offering the advantage of customizable range and resolution for calculations. Comparatively, the discrete-time design of the IRC proves computationally more efficient than its continuous-time counterpart. Observations indicate that the discrete-time IRC requires less memory and computational resources for its implementation within the same data format, thus minimizing the latency.

\subsection{Discrete-time IRC Design}
According to Theorem 1, the closed-loop interconnection of the minimal system (\ref{eq:plant}) and a discrete-time IRC of the form (\ref{eq:DT IRC model}) is asymptotically stable if the discrete-time IRC (\ref{eq:DT IRC model}) satisfies $-2\Gamma^{-1}< D<-G(1)$. Controller parameters can be tuned to achieve the desired performance in the time and frequency domain. Accordingly, $D = -3$ and $\Gamma=0.010$ are opted for in the discrete-time IRC.

\begin{figure}[t]
	\centering
	\includegraphics[width=8.5cm]{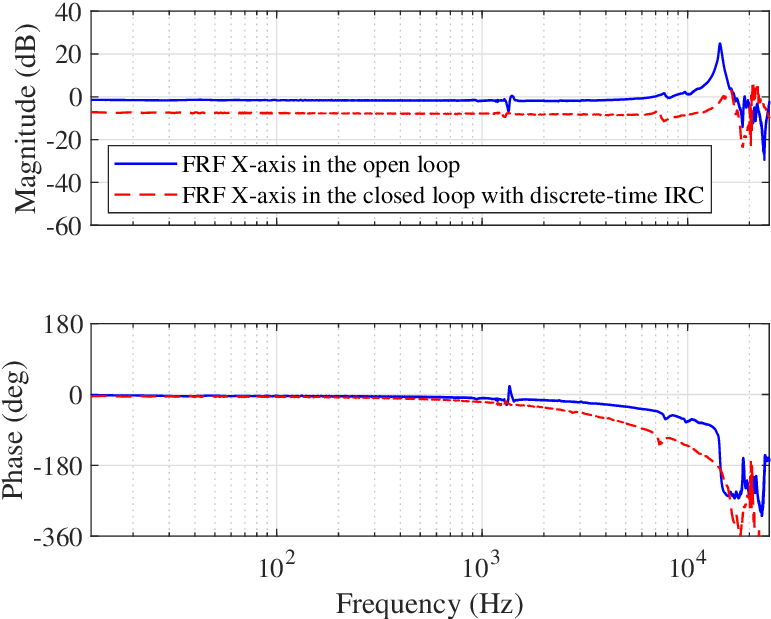}
	\caption{\footnotesize Bode plot of the flexure-guided nanopositioner in open loop
and in a positive feedback interconnection with the discrete-time IRC.}\label{fig:dampedx}
\end{figure}

The frequency response of the nanopositioner in a closed loop with the discrete-time IRC is depicted in Fig.\ref{fig:dampedx}. We observe that a substantial damping of about $14.4\,dB$ is achieved at the resonance.

\begin{figure}[t]
	\centering
	\includegraphics[width=8.5cm]{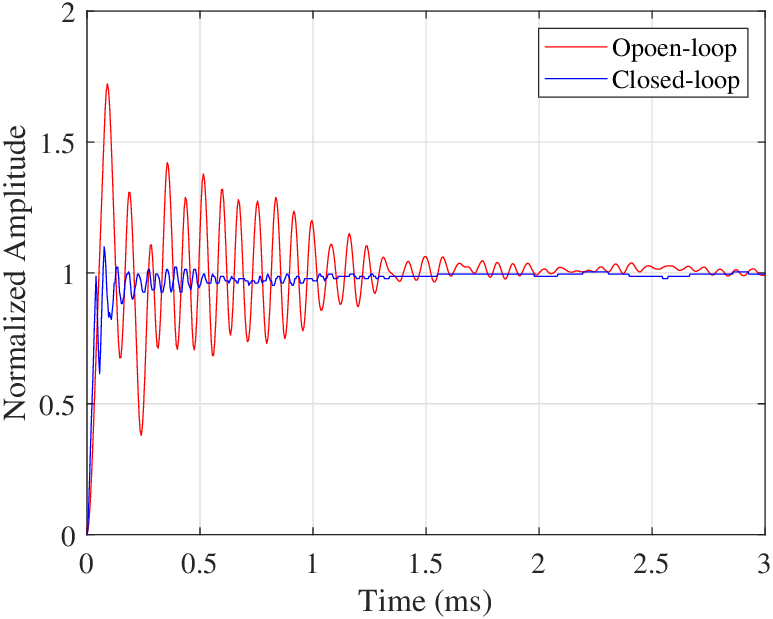}
	\caption{\footnotesize Normalized Step Response of the flexure-guided nanopositioner in open loop and damped closed-loop.}\label{fig:step}
\end{figure}

We also applied a unity step as an input disturbance to the system and analyzed the damped system response in a closed loop. Figure \ref{fig:step} demonstrates the step response of the flexure-guided nanopositioner in an open loop and closed loop with the discrete-time IRC in a positive feedback interconnection.

\section{CONCLUSION}\label{sec:conclusion}
In this article, we introduce the discrete-time IRC to provide efficient and rigorous digital control for NI systems. We show that a discrete-time IRC has SANI property. With suitable parameters chosen, a discrete-time IRC can asymptotically stabilize an NI system. This stability result motivates the application of the discrete-time IRC on a high-speed flexure-guided nanopositioner. It is shown in a hardware experiment that a nanopositioner can be effectively damped by a discrete-time IRC.

\bibliographystyle{IEEEtran}

\end{document}